\documentclass[12pt]{article}
\usepackage{simplemargins, amsmath, array, delarray, emlines2}
\setallmargins{1in}
\def\lbracket{\left\langle\negmedspace\left\langle}
\def\rbracket{\right\rangle\negmedspace\right\rangle}

\begin{document}

\thispagestyle{empty}
\begin{flushright}
hep-th/0105314
\end{flushright}
\vskip 3.5cm

\begin{center}\LARGE
{\bf Descent Relations in Type 0A/0B}
\end{center}
\vskip 1.0cm
\begin{center}
{\large  David Mattoon Thompson\footnote{E-mail: {\tt mattoon@physics.harvard.edu}}}

\vskip 0.5 cm
{\it Jefferson Physical Laboratory\\
Harvard University\\
Cambridge, MA  02138}
\end{center}

\vskip 0.8cm

\begin{center}
May 2001
\end{center}

\vskip 2cm


\begin{abstract}
The type 0 theories have twice as many stable D-branes as the type II theories.  In light
of this added complication, we find the descent relations for D-branes in the type 0A
and 0B theories.  In addition, we work out how the two types of D-branes differ in their couplings to NS-NS fields.
\end{abstract}

\vfill
\newpage
\renewcommand{\baselinestretch}{1.4}


\section*{Introduction}

In this paper, we gain further insight into type 0 D-branes by working out the descent
relations for type 0 theories. Sen's descent relations in the type II theories relate
different D-branes through operations of orbifolding and tachyon kinking.  These relations
form an interlocking chain of relationships between the different types of D-branes.  
Although the type 0 theories are in many ways similar to the type II theories, it is not
immediately clear how one should draw the descent relation diagram since type 0 theories
have twice the number of D-branes.  This problem is addressed in sections 3 through 5.

Sections 1 and 2 serve as very brief introductions to the type 0 theories and their
D-brane content.  In section 3, we review the descent relations in type II theories and we
manage to rule out certain combinations of type 0 D-branes from having any starring role
in the type 0 descent relations.  In sections 4 and 5, we uncover how the type 0 D-branes
are related via orbifolds and kinks, respectively.  By the end of section 5, we have
pieced together the type 0 descent relations.

Section 6 demonstrates the fundamental distinction between the two types of D-branes in
type 0 theories.  We show in this last section that the two types of D-branes, D+ branes
and D$-$ branes, have opposite charges with respect to all (NS$-$,NS$-$) fields.  We will 
also show how a general disk amplitude with a D+ relates to the same amplitude with a 
D$-$.


\section{Perturbative Spectrum}

Type II superstring theories are composed of left and right moving pieces which reside in
one of four sectors, NS$\pm$ and R$\pm$.  The + and $-$ here denote the value of the
worldsheet fermion number operator, $(-1)^{F}$, not to be confused with the $(-1)^{F_L^s}$
operator to be introduced later.  At first blush, it appears as though there are on the
order of $2^{16}$ possible string theories, each factor of 2 coming from whether or
not a given theory contains a particular combination of sectors.  Several consistency
conditions pare this enormous number of possibilities to only four.  Two of these are the
type IIA and IIB theories.  The other two are the less familiar type 0A and 0B theories.

The consistency conditions are as follows (for a review, see \cite{polchII}):
\begin{description}
\item{\textbf{Level matching}:}
The first condition we use to rule out some theories is the level matching condition $L_0 
= \tilde{L}_0$.  The NS$-$ sector has half-integer levels while the NS+, R+, and R$-$ 
have integer levels.  Therefore, NS$-$ can not be paired with any of the other three sectors.  
\item{\textbf{Mutual locality}:}
All pairs of vertex operators must be mutually local.  That is, the phase obtained by
taking one vertex operator in a circle around the other must be unity or else there is
phase ambiguity in the amplitude.
\item{\textbf{Closed OPE}:}
The OPE of the vertex operators in the theory must be in terms of vertex operators that
are also present in the theory.
\item{\textbf{Modular invariance}:}
Modular invariance requires that there be at least one left moving R sector and at least 
one right moving R sector.
\end{description}
The only four theories that satisfy these simple consistency requirements are the type IIA 
and IIB theories,
\begin{equation}
\begin{array}{ccccc}
\text{IIA:} & \text{(NS+,NS+)} & \text{(R+,R$-$)} & \text{(NS+,R$-$)} & \text{(R+,NS+)} \\
\text{IIB:} & \text{(NS+,NS+)} & \text{(R+,R+)} & \text{(NS+,R+)} & \text{(R+,NS+)}
\end{array}
\label{IIspec}
\end{equation}
and the type 0A and 0B theories,
\begin{equation}
\begin{array}{ccccc}
\text{0A:}&\text{(NS+,NS+)}&\text{(NS$-$,NS$-$)}&\text{(R+,R$-$)}&\text{(R$-$,R+)}\\
\text{0B:}&\text{(NS+,NS+)}&\text{(NS$-$,NS$-$)}&\text{(R+,R+)}&\text{(R$-$,R$-$)}\, .
\end{array}
\label{0spec}
\end{equation}
The perturbative spectra of the type 0 theories contain no spacetime fermions.  In the 
NS-NS sectors, the low-lying states are the tachyon from (NS$-$,NS$-$) and the graviton, 
antisymmetric tensor, and dilaton from (NS$+$,NS$+$).  The type 0 theories have twice as 
many massless R-R states as the type II theories.  In particular, type 0A has two 
R-R 1-forms and two R-R 3-forms; type 0B has two R-R scalars, two R-R 2-forms, and one R-R 
4-form with an unconstrained 5-form field strength.


\section{D-branes}

The fact that the type 0 theories have twice as many R-R fields as the type II theories is
an indication that there may be twice as many stable D-branes in type 0 as compared to
type II.  This turns out to be correct and can be understood quite directly by examining
D-branes in the boundary state formalism (for a review, see \cite{0005029}).  In this
formalism, D-branes are represented by boundary states for the physical closed strings.  
These boundary states are themselves coherent closed string states. 

In both the type II and type 0 theories, there are four types of boundary states for
each~p, 
\begin{equation} 
|Bp,+\rangle_{\text{NS-NS}}\, , \quad
|Bp,-\rangle_{\text{NS-NS}}\, , \quad |Bp,+\rangle_{\text{R-R}}\, , \quad
|Bp,-\rangle_{\text{R-R}}\, . 
\end{equation} 
The + and $-$ denote the boundary conditions on the worldsheet fermions and superghosts 
as in equations (\ref{dbranebc}).
Linear combinations of these states must be
taken to form D-brane boundary states which, in turn, must be GSO-invariant and must 
satisfy certain consistency conditions \cite{0005029}.

The D-brane boundary states in the type 0 theories are 
as follows:
\begin{equation}
\begin{array}.{rcl}\}
|Dp,+\rangle &=& |Bp,+\rangle_{\text{NS-NS}} + |Bp,+\rangle_{\text{R-R}}\\
|Dp,-\rangle &=& |Bp,-\rangle_{\text{NS-NS}} + |Bp,-\rangle_{\text{R-R}}\\
|\overline{D}p,+\rangle &=& |Bp,+\rangle_{\text{NS-NS}} - |Bp,+\rangle_{\text{R-R}}\\  
|\overline{D}p,-\rangle &=& |Bp,-\rangle_{\text{NS-NS}} - |Bp,-\rangle_{\text{R-R}}
\end{array}
\; \text{for p even (odd) in 0A (0B)}
\label{stabdstates}
\end{equation}
\begin{equation}
\begin{array}.{rcl}\}
|\widehat{Dp},+\rangle &=& |Bp,+\rangle_{\text{NS-NS}}\\
|\widehat{Dp},-\rangle &=& |Bp,-\rangle_{\text{NS-NS}}
\end{array}
\; \text{for p odd (even) in 0A (0B)}.
\label{unstabdstates}
\end{equation}
Using $\eta$ to denote $\pm 1$, the $|Dp, \eta\rangle$ states correspond to stable
D-branes.  We see from the minus sign in front of the R-R boundary states that the
$|\overline{D}p,\eta\rangle$ states correspond to stable anti-D-branes.  The
$|\widehat{Dp}, \eta\rangle$ states correspond to unstable D-branes.

Let us pause for a second to make a remark on D-brane stability.  The condition for
stability is that the spectrum of open strings on the D-brane does not contain a tachyon.  
It is important not to confuse this condition with being BPS.  Of course, none of the
D-branes can be BPS in the type 0 theories since there is no supersymmetry to begin with;
there are no fermions in the absence of D-branes.  It just so happened for D-branes in
the type II theories that the conditions of stability and BPS coincided.

It will be important for our purposes to find the spectra of open strings living on or 
between D-branes.  The details can be found in Appendix A and the results for type 0 
D-branes are given in tables 1 and 2.  The spectra in table 1 can be extrapolated to all 
possibilities by noting that a given spectrum is invariant under the replacements D 
$\leftrightarrow$ $\overline{\text{D}}$ and/or $+\leftrightarrow -$.  For example, from 
the first line of table 1, we see that the open strings beginning on a Dp+ and ending on a 
Dp+ are NS+.  Therefore, the strings beginning on a $\overline{\text{D}}$p+ and ending on 
a $\overline{\text{D}}$p+ are NS+.  Similarly, strings beginning on a Dp$-$ and ending on 
a Dp$-$ (or beginning on a $\overline{\text{D}}$p$-$ and ending on a 
$\overline{\text{D}}$p$-$) are also NS+.
\begin{center}
\begin{tabular}{|c|c|c|}
\hline
\multicolumn{3}{|c|}{Open Spectrum on Stable D-branes}\\
\multicolumn{3}{|c|}{(p odd in 0B, p even in 0A)}\\
\hline\hline
$\sigma=0$ & $\sigma=\pi$ & Spectrum \\
\hline
Dp+ & Dp+ & NS+ \\
\hline
Dp+ & $\overline{\text{D}}$p+ & NS$-$ \\
\hline
Dp+ & Dp$-$ & R+ \\
\hline
Dp+ & $\overline{\text{D}}$p$-$ & R$-$ \\
\hline
\end{tabular}\\
\vspace{5pt}
Table 1: All other cases obtained by one or both of the following \\
operations under which the spectrum is invariant:
$+ \leftrightarrow -$, $\text{D} \leftrightarrow \overline{\text{D}}$.
\end{center}
We see that there are two tachyons among the open strings stretched between a
$|Dp,\eta\rangle$ and a $|\overline{D}p,\eta\rangle$.  One tachyon starts (at $\sigma=0$)
on the $|Dp,\eta\rangle$ and ends (at $\sigma=\pi$) on the $|\overline{D}p, \eta\rangle$,
and the other tachyon starts on the $|\overline{D}p,\eta\rangle$ and ends on the
$|Dp,\eta\rangle$.  This indicates an instability in the D$\overline{\text{D}}$ pair.
\begin{center}
\begin{tabular}{|c|c|c|}
\hline
\multicolumn{3}{|c|}{Open Spectrum on Unstable D-branes}\\
\multicolumn{3}{|c|}{(all p in 0A and 0B)}\\
\hline\hline
$\sigma=0$ & $\sigma=\pi$ & Spectrum \\
\hline
$\widehat{\text{Dp}}$+ & $\widehat{\text{Dp}}$+ & NS+, NS$-$ \\
\hline
$\widehat{\text{Dp}}$+ & $\widehat{\text{Dp}}-$ & R+, R$-$ \\
\hline
\end{tabular}\\
\vspace{5pt}
Table 2: All other cases obtained by
$+ \leftrightarrow -$ \\ under which the spectrum is invariant.
\end{center}
We see in table 2, as expected, that there is a tachyon living on 
the unstable
$|\widehat{Dp},\eta\rangle$ D-branes.


\section{Descent Relations}

Sen's descent relations give relations between different D-brane configurations in the
type II theories (for a review, see \cite{9904207}).  The two important operations are
orbifolding by $(-1)^{F_L^s}$, where $F_L^s$ is the spacetime fermion number of the
left-movers, and kinking the tachyon field that lives on unstable configurations of
D-branes.  Starting with a coincident D(2p)$\overline{\text{D}}$(2p) pair in type IIA,
orbifolding by $(-1)^{F_L^s}$ yields an unstable $\widehat{\text{D(2p)}}$ in type IIB.  
Orbifolding one more
time leaves us with a stable D(2p) in the type IIA theory.  Starting again with the
D(2p)$\overline{\text{D}}$(2p)  pair in type IIA, but this time kinking the tachyon field
that lives on the D-branes, we are left with an unstable $\widehat{\text{D(2p$-$1)}}$ in 
type IIA.  Kinking
the remaining tachyon field gives us a stable D(2p$-$2) in type IIA.  The results are
similar if we start with a D(2p+1)$\overline{\text{D}}$(2p+1) pair in type IIB.  In fact,
the descent relations form an interlocking chain as shown in figure 1.


\begin{center}
\begin{tabular}{c@{}c@{}c@{}c@{}c@{}c@{}c@{}c}
& & & & & $\downarrow$ & & $\downarrow$ \\
& & & IIB D(2p+1)$\overline{\text{D}}$(2p+1) & $\;\rightarrow\;$ & IIA 
$\widehat{\text{D(2p+1)}}$ & 
$\;\rightarrow\;$ & IIB D(2p+1) \\
& & & $\downarrow$ & & $\downarrow$ & & \\
& IIA D(2p)$\overline{\text{D}}$(2p) & $\;\rightarrow\;$ & IIB $\widehat{\text{D(2p)}}$ & 
$\rightarrow$ & 
IIA D(2p) & & \\
& $\downarrow$ & & $\downarrow$ & & & &\\
$\rightarrow$ & IIA $\widehat{\text{D(2p$-$1)}}$ & $\rightarrow$ & IIB D(2p$-$1) & & & 
& 
\\
& $\downarrow$ & & & & & & \\
$\rightarrow$ & IIA D(2p$-$2) & & & & & &
\end{tabular}\\
\vspace{5pt}
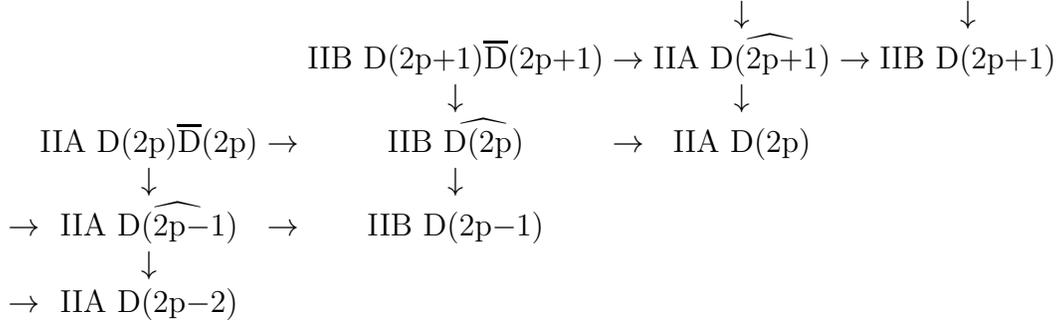
Figure 1: Descent relations for the type II theories.  Horizontal arrows \\ 
denote modding by $(-1)^{F_L^s}$.  Vertical arrows denote the tachyonic kink.
\end{center}

The natural question at this point is what the analogue of the descent relations is for
the type 0 theories.  Starting with a D(2p)$\overline{\text{D}}$(2p) in type 0A, we have
four possibilities to consider: a choice of + or $-$ for each of the two branes.  Then,
once we orbifold (kink), we must figure out whether we get $\widehat{\text{D(2p)}}$+ or
$\widehat{\text{D(2p)}}$$-$ ($\widehat{\text{D(2p$-$1)}}$+ or
$\widehat{\text{D(2p$-$1)}}-$).  For a discussion of the differences between D+ and D$-$
branes, see section 6.

In the type II descent relations, every time we orbifold or kink we effectively remove
one of the tachyonic degrees of freedom.  A complex tachyon lives on the
D$\overline{\text{D}}$ pair; orbifolding or kinking once gives an unstable D-brane with a
real tachyon; orbifolding or kinking one more time gives a stable D-brane with no tachyon
field.  With this observation, we can quickly rule out two of the choices for the
D$\overline{\text{D}}$ pair in the type 0 case.  Since the open string tachyon arises
from the NS$-$ sector, we see from table 1 that only the Dp+$\overline{\text{D}}$p+ and
Dp$-\overline{\text{D}}$p$-$ pairs for p odd in 0B (even in 0A) have tachyon fields
living on them.

Holding out some hope for the Dp+$\overline{\text{D}}$p$-$ pair, let us see if there 
is any room in the type 0 descent relations for this object.  Clearly, we can not consider 
a tachyon kink since there is no tachyonic kink on this pair of D-branes: from table 1, 
we see that there are NS+ strings living on each of the D-branes and R$-$ strings 
stretched between the two.  Perhaps we can orbifold this pair of D-branes by 
$(-1)^{F_L^s}$.  However, one can take the $(-1)^{F_L^s}$ orbifold in the presence of 
D-branes only if that configuration of D-branes is invariant under $(-1)^{F_L^s}$. 
For example, in the type II theories, 
$(-1)^{F_L^s}|D(2p)\rangle=|\overline{D}(2p)\rangle$ and 
$(-1)^{F_L^s}|\overline{D}(2p)\rangle=|D(2p)\rangle$, so we were able to orbifold the 
D$\overline{\text{D}}$ pairs.  
Since 
\begin{eqnarray}
(-1)^{F_L^s}|Bp,\pm\rangle_{\text{NS-NS}} &=& |Bp,\pm\rangle_{\text{NS-NS}}\, , 
\nonumber \\
(-1)^{F_L^s}|Bp,\pm\rangle_{\text{R-R}} &=& -|Bp,\pm\rangle_{\text{R-R}}\, ,
\end{eqnarray}
we see from (\ref{stabdstates}) that in the type 0 theories
\begin{eqnarray}
(-1)^{F_L^s}|Dp+\rangle &=& |\overline{D}p+\rangle\, , \nonumber \\
(-1)^{F_L^s}|Dp-\rangle &=& |\overline{D}p-\rangle\, ,
\label{switch1}
\end{eqnarray}
and
\begin{eqnarray}
(-1)^{F_L^s}|\overline{D}p+\rangle &=& |Dp+\rangle\, , \nonumber \\
(-1)^{F_L^s}|\overline{D}p-\rangle &=& |Dp-\rangle\, .
\label{switch2}
\end{eqnarray}
This means that the coincident Dp+$\overline{\text{D}}$p$-$ pair is not invariant under 
$(-1)^{F_L^s}$ and we no longer consider it as a potential participant in the type 0 
descent relations.  Fortunately, the Dp+$\overline{\text{D}}$p+ and
Dp$-\overline{\text{D}}$p$-$ pairs \textit{are} invariant under $(-1)^{F_L^s}$, so we 
will be able to interpret the orbifold as a projection of the open string states.


\section{$(-1)^{F_L^s}$ Orbifold}

Here we will consider what happens to the coincident D(2p)+$\overline{\text{D}}$(2p)+ pair
in type 0A under the $(-1)^{F_L^s}$ orbifold.  First, let us look at the spacetime bulk
far from the D-branes.  Locally, this is just type 0A without any open strings.  Taking
the orbifold of type 0A by $(-1)^{F_L^s}$ gives the type 0B theory, and vice versa (see 
Appendix B for details).

As we have already noted in equations (\ref{switch1}) and (\ref{switch2}) , $(-1)^{F_L^s}$ 
switches the D(2p)+ and 
$\overline{\text{D}}$(2p)+, so its action on the Chan-Paton factors is
\begin{equation}
\Lambda \rightarrow \sigma_1 \Lambda \sigma_1^{-1}\, .
\end{equation}
Of the four Chan-Paton factors, $I$, $\sigma_1$, $\sigma_2$, and $\sigma_3$, only $I$ and 
$\sigma_1$ are invariant under this operation.  Therefore, the open strings with CP 
factors $I$ and $\sigma_1$ are kept and those with CP factors $\sigma_2$ and $\sigma_3$ 
are thrown out.

We can see that this new object, the result of orbifolding
D(2p)+$\overline{\text{D}}$(2p)+, is a single brane since the degrees of freedom
corresponding to the relative positions of the original D-branes have been projected out.  
The position coordinates corresponding to their respective CP factors are as given below.
\\
\begin{minipage}{6.7cm}
\begin{picture}(6.7, 3)
\put(2.7,.1){\line(0,1){2}}
\put(3.7,.6){\line(0,1){2}}
\put(4.7,.1){\line(0,1){2}}
\put(5.7,.6){\line(0,1){2}}
\put(2.7,.1){\line(2,1){1}}
\put(2.7,2.1){\line(2,1){1}}
\put(4.7,.1){\line(2,1){1}}
\put(4.7,2.1){\line(2,1){1}}
\put(3.1,0){$x_0$}
\put(5.1,0){$y_0$}
\put(3.1,2.5){1}
\put(5.1,2.5){2}
\end{picture}
\end{minipage}
\begin{minipage}{7cm}
\begin{eqnarray*}
1-1 &:& X=x_0 + \ldots \\
2-2 &:& X=y_0 + \ldots \\
1-2 &:& X=x_0 + \frac{\sigma}{\pi} (y_0-x_0) + \ldots\\
2-1 &:& X=y_0 + \frac{\sigma}{\pi} (x_0-y_0) + \ldots
\end{eqnarray*}
\end{minipage}
\vspace{3mm} \\
Writing out the lowest order degrees of freedom in terms of Chan-Paton factors, we find 
that we can regroup them as
\begin{multline}
x_0 \begin{pmatrix} 1 & 0 \\ 0 & 0 \end{pmatrix} + 
y_0 \begin{pmatrix} 0 & 0 \\ 0 & 1 \end{pmatrix} + 
[x_0 + \frac{\sigma}{\pi} (y_0 - x_0)] \begin{pmatrix} 0 & 1 \\ 0 & 0 \end{pmatrix} +
[y_0 + \frac{\sigma}{\pi} (x_0 - y_0)] \begin{pmatrix} 0 & 0 \\ 1 & 0 \end{pmatrix} \\
= \frac{1}{2} (x_0 + y_0) I + \frac{1}{2}(x_0-y_0) \sigma_3 + \frac{1}{2}(x_0+y_0) 
\sigma_1 + \frac{1}{2} [-i (x_0-y_0) + \frac{2i\sigma}{\pi} (x_0-y_0)]\sigma_2\, .
\end{multline}
The $(x_0-y_0)$ degree of freedom multiplies only $\sigma_2$ and $\sigma_3$, which are 
projected out.

After orbifolding, we are left with a (2p)-brane in the type 0B theory
with NS+ strings (corresponding to I) and NS$-$ strings (corresponding to
$\sigma_1$) living on it.  This identifies the object as either
$\widehat{\text{D(2p)}}+$ or $\widehat{\text{D(2p)}}-$.  In order to
distinguish between these two options, we look at the coupling of this
(2p)-brane to the (NS$-$,NS$-$) tachyon and compare it to the coupling of
the $\widehat{\text{D(2p)}}+$ and $\widehat{\text{D(2p)}}-$ to the
(NS$-$,NS$-$) tachyon.  But first we must determine what these couplings
are.

We know from \cite{9811035} that \textit{stable} D-branes in the type 0 theories have the 
term
\begin{equation}
-\frac{T_p q \overline{q}}{4} \int d^{p+1}\sigma \, T(X) 
\label{effcoup1}
\end{equation}
in their low energy effective action, where $T$ is the closed string tachyon, and $q$ and 
$\overline{q}$ are the D-brane's charges under the massless R-R fields $C$ and 
$\overline{C}$.  The R-R charges of stable D-branes in the type 0 theories are given in 
table 3.  Notice that $q\overline{q}=\eta$.
\\  
\begin{center}
\begin{tabular}{|c|c|c|}
\hline
\multicolumn{3}{|c|}{Stable Dp R-R Charges}\\
\multicolumn{3}{|c|}{(p odd in 0B, p even in 0A)}\\
\hline\hline
$\qquad\qquad\qquad$ & q & $\overline{q}$ \\
\hline
Dp+ & 1 & 1 \\
\hline
$\overline{\text{D}}$p+ & $-$1 & $-$1 \\
\hline
Dp$-$ & 1 & $-$1 \\
\hline
$\overline{\text{D}}$p$-$ & $-$1 & 1 \\
\hline
\end{tabular}\\
\vspace{5pt}
Table 3
\end{center}
\vspace{3mm}
We know from cylinder diagrams between D-branes that the \textit{unstable} 
$\widehat{\text{D+}}$ 
and $\widehat{\text{D$-$}}$ have opposite tachyon charge \cite{9908121}, but this can not 
tell us how to 
assign 
the charges to the two types of D-branes.  The solution to this can be found by comparing 
tachyon tadpole calculations for the stable and unstable D-branes.

The amplitude \cite{9802088, 0004198} for a stable Dp+ to emit a tachyon is
\begin{eqnarray}
\langle T,k | Dp,+\rangle &=& \langle T,k | (|Bp,+\rangle_{\text{NS-NS}} + |Bp,+\rangle_{ 
\text{R-R}}) \nonumber \\
&=& \langle T,k | Bp, +\rangle_{\text{NS-NS}} \nonumber \\
&=& \langle e^{-\Phi - \tilde{\Phi}} e^{-ik\cdot X} |Bp, 
+\rangle_{\text{NS-NS}}\nonumber\\
&=& \frac{T_p}{2}\langle e^{-\Phi -\tilde{\Phi}} e^{-ik\cdot X} |B_X\rangle 
|B_{\text{gh}}\rangle |B_\psi, \eta\rangle_{\text{NS-NS}} |B_{\text{sgh}}, 
\eta\rangle_{\text{NS-NS}}\, ,
\label{tachamp1}
\end{eqnarray}
where $k$ is perpendicular to the D-brane.  Now consider an unstable 
$\widehat{\text{D(p$-$1)}}$+ that is extended in p$-$1 of the same 
directions as the Dp+.
The amplitude for an unstable $\widehat{\text{D(p$-$1)}}$+ to emit a tachyon in the same 
direction is
\begin{eqnarray}
\langle T,k|\widehat{D(p-1)},+\rangle &=& \langle T,k | 
B(p-1),+\rangle_{\text{NS-NS}} 
\nonumber\\
&=& \frac{T_{p-1}}{2}\langle e^{-\Phi -\tilde{\Phi}} e^{-ik\cdot X} |B_X\rangle^\prime
|B_{\text{gh}}\rangle |B_\psi, \eta\rangle_{\text{NS-NS}}^\prime 
|B_{\text{sgh}},
\eta\rangle_{\text{NS-NS}}\, .
\label{tachamp2}
\end{eqnarray}
The only difference between (\ref{tachamp1}) and (\ref{tachamp2}) is the normalization and
the matter part of the boundary state.  Both $T_p$ and $T_{p-1}$ are positive constants.
The difference between $|B_X\rangle^\prime$ and $|B_X\rangle$ is a
minus sign on one of the $X$ fields which does not get contracted with the $e^{ik\cdot X}$
of the tachyon since $k$ is perpendicular to the Dp+.  The 
difference between $|B_\psi\rangle^\prime$ and $|B_\psi\rangle$ is a minus sign on one of 
the $\psi$ fields, but none of the $\psi$ fields in the boundary state get contracted with 
anything in the tachyon vertex operator.  Therefore, the tachyon charge of the unstable 
$\widehat{\text{D(p$-$1)}}$+ is related to the charge of the stable Dp+ by a factor of 
$T_{p-1}/T_p$, so the 
tachyon tadpole term in an unstable $|\widehat{D(p-1)},\eta\rangle$ brane's low energy 
effective 
action is
\begin{equation}
-\frac{T_{p-1} \eta}{4} \int d^{p+1}\sigma \, T(X)\, .
\label{effcoup2}
\end{equation} 
Note, by comparing (\ref{effcoup1}) and (\ref{effcoup2}), that the Dp+ and the 
$\widehat{\text{D(p$-$1)}}$+ 
couple with the same sign to the closed string tachyon.

Since both the closed string tachyon and the NS-NS boundary state part of the D-branes
both reside in the (NS,NS) sector which is unaffected by the orbifold, the coupling of
the brane to the tachyon should be unchanged.  This means that the
D(2p)+$\overline{\text{D}}$(2p)+ in type 0A gets orbifolded to the
$\widehat{\text{D(2p)}}$+ of the type 0B.

We can understand the orbifold at the level of boundary states by considering the
emission and reabsorption of closed strings by the D(2p)+$\overline{\text{D}}$(2p)+
pair.  To simplify our equations, we introduce the shorthand notation
\begin{equation}
\lbracket\Lambda\rbracket \equiv \int dl\, 
\begin{pmatrix}|D(2p)+\rangle\\ |\overline{D}(2p)+\rangle \end{pmatrix}^\dagger\,
e^{-lH_c} \Lambda\, 
\begin{pmatrix}|D(2p)+\rangle\\ |\overline{D}(2p)+\rangle \end{pmatrix}\, .
\end{equation}
In this formalism, the calculation of the cylinder diagram for an open string with CP
factor $\Lambda$ can be rewritten as the closed string exchange amplitude $\lbracket
\Lambda \rbracket$.
The amplitude for a closed string to be emitted and reabsorbed by the 
D(2p)+$\overline{\text{D}}$(2p)+ pair is equal to
\begin{equation}
\lbracket\begin{pmatrix}1&1\\1&1\end{pmatrix}\rbracket = \lbracket I + \sigma_1 
\rbracket\, .
\label{exch1}
\end{equation}
When we orbifold by projecting out $\sigma_2$ and $\sigma_3$, we see that this
amplitude is unchanged.  However, we know from our earlier discussion that the 
resulting object is a single D-brane.  Therefore, we should be able to rewrite 
(\ref{exch1}) as the emission and absorption of a closed string by a single 
$\widehat{\text{D(2p)}}$.  Attempting this, we find
\begin{equation}
\lbracket I + \sigma_1 \rbracket \quad = \quad
\begin{cases}
4 \int dl\, \langle \widehat{D(2p)}+|e^{-lH_c}|\widehat{D(2p)}+\rangle\\
4 \int dl\, \langle \widehat{D(2p)}-|e^{-lH_c}|\widehat{D(2p)}-\rangle\, .
\end{cases}
\end{equation}
This amplitude can be written in terms of either a $\widehat{\text{D(2p)}}$+ or a
$\widehat{\text{D(2p)}}-$, but our previous tachyon charge argument singles out the
$\widehat{\text{D(2p)}}$+.

If we orbifold one more time by $(-1)^{F_L^s}$, the bulk transforms back to type 0A.  The
action of the orbifold on the D-brane's open string modes can be determined by examining
the two-point functions of the theory.  The existence of nonzero two-point functions
between open strings on the D-brane and closed strings in the bulk allows us to determine
the action of $(-1)^{F_L^s}$ on the open strings by requiring the correlation functions to
be invariant.  As in the type II case \cite{9904207}, the orbifold's effect on the
$\widehat{\text{D(2p)}}$+ is to project out the open strings with CP factor $\sigma_1$.  
Removing the $\sigma_1$ from (\ref{exch1}) leaves the following amplitude for closed
string emission and absorption:
\begin{equation}
\lbracket I \rbracket \quad = \quad
\begin{cases}
2 \int dl\, \langle D(2p)+ | e^{-lH_c} |D(2p)+\rangle \nonumber \\
2 \int dl\, \langle \overline{D}(2p)+ | e^{-lH_c} |\overline{D}(2p)+\rangle 
\nonumber \\
2 \int dl\, \langle D(2p)- | e^{-lH_c} |D(2p)-\rangle \nonumber \\
2 \int dl\, \langle \overline{D}(2p)- | e^{-lH_c} |\overline{D}(2p)-\rangle 
\, .
\end{cases}
\end{equation}
This time, the amplitude can be written in four ways, in terms of a D(2p)+,
$\overline{\text{D}}$(2p)+, D(2p)$-$, or $\overline{\text{D}}$(2p)$-$.  Based on the
previous tachyon charge argument, we can rule out the last two possibilities, so we know
the resulting object is either a stable D(2p)+ or a stable $\overline{\text{D}}$(2p)+
in type 0A.  This agrees with Sen's observation in \cite{9904207} that there is an
inherent ambiguity as to whether the resulting object is a brane or an anti-brane.


\section{Tachyonic Kink}

The other component to the descent relations is the tachyonic kink.  As shown in figure
1, kinking one of the two tachyons on a Dp$\overline{\text{D}}$p in a type II theory
yields a $\widehat{\text{D(p$-$1)}}$ in the same theory and kinking the remaining tachyon
results in a D(p$-$2).  This part of the descent relations is shown by taking a series of
marginal deformations that connect the Dp$\overline{\text{D}}$p to the tachyonic kink and
following what happens to the CFT under these deformations.

To outline the series of marginal deformations, we will use the D1$\overline{\text{D}}$1 
pair in 0B for simplicity.  The details of this analysis can be found in  \cite{9904207, 
9808141}.  We begin 
with the D1$\overline{\text{D}}$1 pair wrapped on a circle 
of radius $R$ and make the following deformations:
\begin{enumerate}
\item We increase the gauge field on the $\overline{\text{D}}$1 so that the open strings 
with CP factors $\sigma_1$ and $\sigma_2$ are antiperiodic around the compactification 
circle.  In particular, the tachyon field with CP factor $\sigma_1$ is moded by 
half-integers as
\begin{equation}
T(x,t)=\sum\limits_{n\in \mathbf{Z}}T_{n+\frac{1}{2}}(t)e^{i(n+\frac{1}{2})\frac{x}{R}}\, 
.
\end{equation}
\item The radius of the circle is taken down to $R = 1/\sqrt{2}$.  At this value, the 
$T_{\pm\frac{1}{2}}$ modes are massless and, therefore, correspond to marginal 
deformations.
\item A vev of $-i$ is given to $(T_{\frac{1}{2}} - T_{-\frac{1}{2}})$ which 
corresponds to
\begin{equation}
T(x) = \sin \frac{x}{2R}\, .
\end{equation}
This is the tachyonic kink.
\item The radius, $R$, is taken back to infinity.
\end{enumerate}

Step number three will be our main focus.  In order to understand the effect of this step,
we first bosonize the worldsheet spinors $\psi_L$ and $\psi_R$ (often denoted as $\psi$
and $\tilde{\psi}$)  whose spacetime indices correspond to the compactified direction.  
In addition to $\psi_L$, $\psi_R$, and the corresponding $X$ ($=X_L+X_R$), we introduce
four new spinors $\xi_L$, $\xi_R$, $\eta_L$, and $\eta_R$, and two new bosons, $\phi$
($=\phi_L+\phi_R$)  and $\phi^\prime$ ($=\phi^\prime_L + \phi^\prime_R$).  The
bosonization equations relating them are
\begin{eqnarray}
e^{\pm i\sqrt{2}X_L} \sim (\xi_L \pm i\eta_L)\, , \\
e^{\pm i\sqrt{2}\phi_L} \sim (\xi_L \pm i \psi_L)\, , \\
e^{\pm i\sqrt{2}\phi^\prime_L} \sim (\eta_L \pm i \psi_L)\, ,
\end{eqnarray}
and similarly for the right-moving fields.  
We also have the relations \begin{equation}
\xi_L \eta_L \sim \partial X_L\, , \quad \xi_L \psi_L \sim \partial \phi_L \, , \quad 
\eta_L \psi_L \sim \partial \phi^\prime_L \, ,
\end{equation}
as well as the corresponding right-moving relations.
Remember, these fields are specifically those fields 
whose spacetime indices correspond to the compactified direction.  
Written in terms of the new bosonic field, the tachyonic kink is made by inserting
\begin{equation}
\exp\left ( i \frac{\sigma_1}{2\sqrt{2}} \oint \partial \phi\right )
\label{kinkop}
\end{equation}
at the boundary of the disk.
In step four, the radius is taken back to infinity by inserting vertex operators of the 
form $\partial X \overline{\partial}X$.  When the contour integral of $\partial 
\phi$ is contracted around each of these operators, they are converted into $-\partial 
\phi^\prime \overline{\partial}\phi^\prime$.  This corresponds to decreasing the 
$\phi^\prime$ radius, so we must introduce a T-dual variable, $\phi^{\prime\prime}$ 
related to the $\phi^\prime$ as
\begin{equation}
\phi^{\prime\prime}_L = \phi^\prime_L\, , \quad \phi^{\prime\prime}_R = -\phi^\prime_R\, , 
\quad R_{\phi^{\prime\prime}} = 1/R_{\phi^\prime}\, .
\end{equation}
This converts the Neumann boundary condition on $\phi^\prime$ to a Dirichlet boundary 
condition on $\phi^{\prime\prime}$ and we are left with a D0-brane where 
$\phi^{\prime\prime}$ is the new spacetime coordinate in place of $X$.

This process is easily extended to Dp$\overline{\text{D}}$p pairs for $p$ other than 1 
since the other worldsheet fields are left unchanged.  This is, in fact, the key to 
understanding whether a Dp+$\overline{\text{D}}$p+ gets kinked to a 
$\widehat{\text{D(p$-$1)}}$+ or a 
$\widehat{\text{D(p$-$1)}}-$.  Let us take a look now at what the + and $-$ correspond to 
in 
terms of 
worldsheet fields.  The boundary state $|Dp, \eta\rangle$ satisfies the following 
equations:
\begin{eqnarray}
\partial_n X^\mu |Dp,\eta\rangle &=& 0\, , \quad \mu=0,\ldots,p \nonumber\\
(X^i-y^i) |Dp, \eta\rangle &=& 0\, , \quad i=p+1,\ldots,9\nonumber \\
(\psi^\mu - \eta\tilde{\psi}^\mu) |Dp,\eta\rangle&=&0\, , \quad \mu=0,\ldots,p\nonumber \\
(\psi^i + \eta\tilde{\psi}^i) |Dp,\eta\rangle&=&0\, , \quad i=p+1,\ldots,9 \nonumber\\
(b-\tilde{b}) |Dp,\eta\rangle&=&0\, ,\nonumber \\
(c-\tilde{c}) |Dp,\eta\rangle&=&0\, , \nonumber\\   
(\gamma-\eta\tilde{\gamma}) |Dp,\eta\rangle&=&0\, ,\nonumber \\  
(\beta-\eta\tilde{\beta}) |Dp,\eta\rangle&=&0\, .
\label{dbranebc}
\end{eqnarray}
The first four of these equations are the familiar boundary conditions on the matter 
fields.  The last four can be obtained by demanding BRST invariance of the boundary 
state \cite{0004198}.

The only worldsheet fields that are affected by the kink are those whose spacetime index
is the same as the compactified direction.  For example, no matter what tachyonic kinking
procedure we can imagine, $\psi^0$ will certainly be unaffected.  Since the $\eta$ value
of the $|Dp,\eta\rangle$ D-brane can be read off from the boundary condition on $\psi^0$,
$\eta$ is invariant under all marginal deformations corresponding to tachyonic kinks.  
This means that a Dp+$\overline{\text{D}}$p+ gets kinked to a 
$\widehat{\text{D(p$-$1)}}$+.

Now we claim that the rest of the kink analysis goes through the same as it did in the
case of the type II theories.  How can we be so sure of this?  The type 0 and type II
theories differ in their perturbative closed string spectra, but the marginal deformations
needed to bring about a tachyonic kink uses only those parts of the closed string spectra
that type 0 and type II have in common.  In particular, the only closed string vev
that is deformed is that of the graviton which can be found in the (NS+,NS+) sector of all
type 0 and type II theories.  All other deformations have to do with open strings, and the
bosonic open string spectra on D-branes in type 0 and type II theories are identical.  
This can be seen by comparing tables 1 and 2 with tables 4 and 5 in Appendix A.

Let us check that the Dp+$\overline{\text{D}}$p+ gets kinked to the 
$\widehat{\text{D(p$-$1)}}$+ by 
considering the amplitude for the emission of a closed string 
tachyon.  From table 3 and equation (\ref{effcoup1}), we see that the combined 
D1+$\overline{\text{D}}$1+ pair in type 
0B has a nonzero tachyon charge (Recall that $\eta=q\overline{q}$).  The amplitude 
under consideration is the closed tachyon tadpole amplitude: a disk with the tachyon 
vertex operator inserted in the bulk.  Again, kinematics force the momentum of the emitted 
tachyon to be perpendicular to the D1+$\overline{\text{D}}$1+ pair.  Therefore, there are 
no potential contractions between the tachyon vertex operator, $e^{-\Phi-\tilde{\Phi}} 
e^{-ik\cdot X}$, and the tachyonic kink operator in (\ref{kinkop}).  The sign of the 
amplitude is not changed by the marginal deformations, so the result is a 
$\widehat{\text{D0}}$ brane that 
couples to the closed tachyon with the same sign as the D1+$\overline{\text{D}}$1+, namely 
a $\widehat{\text{D0}}$+.

The result we have established here for the D1+$\overline{\text{D}}$1+ pair in type 0B can
easily be extended to all Dp+$\overline{\text{D}}$p+ pairs and
Dp$-\overline{\text{D}}$p$-$ pairs for p even in 0A and p odd in 0B.  The tachyonic kink 
on an unstable $\widehat{\text{Dp}}$+ or $\widehat{\text{Dp}}-$, for $\text{p} >0$, can 
be analyzed by the following 
procedure \cite{9904207}.  Take 
the unstable $\widehat{\text{D1}}$+ in 0A as an example.  If we T-dualize the 
D1+$\overline{\text{D}}$1+ pair 
in type 0B, we find that the D0+$\overline{\text{D}}$0+ pair in 0A is connected by 
marginal deformations to the $\widehat{\text{D1}}$+ in 0A.  By running the marginal 
deformations backwards, 
we see 
that the  D0+$\overline{\text{D}}$0+ corresponds to a kink-antikink pair on the 
$\widehat{\text{D1}}$+.  This 
allows us to identify the tachyonic kink on the $\widehat{\text{D1}}$+ as a stable D0+ in 
type 0A.  The 
flowchart of descent relations in the type 0 theories is given in figure 2.


\begin{center}
\begin{tabular}{c@{}c@{}c@{}c@{}c@{}c@{}c@{}c}
& & & & & $\downarrow$ & & $\downarrow$ \\
& & & 0B D(2p+1)+$\overline{\text{D}}$(2p+1)+ & $\;\rightarrow\;$ & 0A 
$\widehat{\text{D(2p+1)}}$+ & 
$\;\rightarrow\;$ & 0B D(2p+1)+ \\
& & & $\downarrow$ & & $\downarrow$ & & \\
& 0A D(2p)+$\overline{\text{D}}$(2p)+ & $\;\rightarrow\;$ & 0B $\widehat{\text{D(2p)}}$+ & 
$\rightarrow$ & 
0A D(2p)+ & & \\
& $\downarrow$ & & $\downarrow$ & & & &\\
$\rightarrow$ & 0A $\widehat{\text{D(2p$-$1)}}$+ & $\rightarrow$ & 0B D(2p$-$1)+ & & & & 
\\
& $\downarrow$ & & & & & & \\
$\rightarrow$ & 0A D(2p$-$2)+ & & & & & &
\end{tabular}\\*
\vspace{5pt}
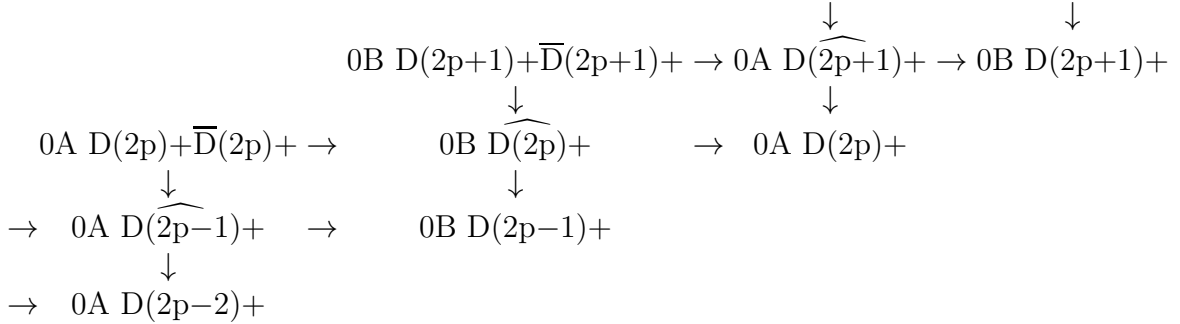
Figure 2: Descent relations for the type 0 theories.  Horizontal arrows \\* 
denote modding by $(-1)^{F_L^s}$.  Vertical arrows denote the tachyonic kink.\\*
A similar diagram exists with $+ \rightarrow -$.
\end{center}


\section{$|Dp,\eta\rangle$: $\eta=+1$ vs. $\eta=-1$}

It is important to stress that the value of $\eta$ in $|Dp,\eta\rangle$ does not just 
affect the R-R charges of the D-brane.  It has an important effect on many string 
amplitudes.  In fact, we will be able to show below that Dp+ and Dp$-$ branes have 
the same tadpole couplings to all (NS+,NS+) fields and opposite tadpole couplings to all 
(NS$-$,NS$-$) fields.

Let us first try to see the opposite tachyon charges of the Dp+ and Dp$-$ at the level of a 
string calculation.  Emission of a tachyon from a D-brane in a type 0 theory is given by a 
disk amplitude with the tachyon vertex operator in the bulk and appropriate boundary conditions 
on the edge.  Note from (\ref{dbranebc}) that these boundary conditions depend 
on $\eta$.  Equations (\ref{dbranebc}) are in terms of the fields defined on the 
upper half plane, so once we map our tachyon amplitude to the upper half plane, the 
following $\eta$-dependent equations must hold on the real axis:
\begin{alignat}{2}
\tilde{\psi}^\mu & = \eta \psi^\mu\, , & \qquad \tilde{\psi}^i & = -\eta \psi^i\, , \\
\tilde{\gamma} & = \eta \gamma\, , & \qquad \tilde{\beta} & = \eta \beta\, .
\end{alignat}
The doubling trick \cite{9611214} extends the string calculation to the entire complex 
plane by defining
\begin{alignat}{2}
\tilde{\psi}^\mu(\overline{z}) & = \eta \psi^\mu(\overline{z})\, , & \qquad
\tilde{\psi}^i(\overline{z}) & = -\eta \psi^i(\overline{z})\, , \\
\tilde{\gamma}(\overline{z}) & = \eta \gamma(\overline{z})\, , & \qquad
\tilde{\beta}(\overline{z}) & = \eta \beta(\overline{z}) 
\end{alignat}
on the lower half plane.
In actual calculations, $\beta$ and $\gamma$ are rebosonized in terms of the free bosons 
$\Phi$ and $\chi$ as
\begin{equation}
\beta \cong e^{-\Phi + \chi} \partial \chi\, , \qquad \gamma \cong e^{\Phi-\chi}\, .
\end{equation}
The doubling trick identifications on $\gamma$ and $\beta$ can be rewritten as 
\begin{eqnarray}
\tilde{\Phi}(\overline{z}) &=& \Phi(\overline{z}) + \frac{i \pi}{2}(1-\eta)\, , 
\nonumber\\
\tilde{\chi}(\overline{z}) &=& \chi(\overline{z})\, .
\end{eqnarray}
After mapping to the upper half plane and then using the doubling trick, the amplitude has 
become
\begin{equation}
\langle e^{-\Phi(z)-\Phi(\overline{z}) - i\pi (1-\eta)/2}e^{-ik\cdot X}\rangle
=(-1)^{(1-\eta)/2} \langle e^{-\Phi(z)-\Phi(\overline{z})} e^{-ik\cdot X}\rangle\, .
\end{equation}
Here we see the explicit dependence on $\eta$ of the D-brane's tachyon charge.

A somewhat complicated, but instructive, example is to look at $C$ goes to $\overline{C}$ 
scattering as depicted in figure 3, where $C$ and $\overline{C}$ are massless bosons from 
the two different R-R sectors.
\begin{center}
\special{em:linewidth 0.4pt}
\unitlength 1.00mm
\linethickness{0.4pt}
\begin{picture}(66.00,51.67)
\emline{1.67}{1.67}{1}{1.67}{41.67}{2}
\emline{1.67}{41.67}{3}{16.67}{51.67}{4}
\emline{16.67}{51.67}{5}{16.67}{11.67}{6}
\emline{16.67}{11.67}{7}{1.67}{1.67}{8}
\put(4.67,7.34){\makebox(0,0)[cc]{Dp}}
\emline{9.67}{26.67}{9}{11.03}{28.17}{10}
\emline{11.03}{28.17}{11}{13.86}{28.17}{12}
\emline{13.86}{28.17}{13}{15.34}{26.67}{14}
\emline{15.34}{26.67}{15}{16.63}{25.07}{16}
\emline{16.63}{25.07}{17}{17.93}{24.69}{18}
\emline{17.93}{24.69}{19}{20.00}{26.67}{20}
\emline{39.00}{26.67}{21}{40.08}{28.31}{22}
\emline{40.08}{28.31}{23}{41.19}{28.99}{24}
\emline{41.19}{28.99}{25}{42.34}{28.70}{26}
\emline{42.34}{28.70}{27}{44.00}{26.67}{28}
\emline{44.00}{26.67}{29}{46.40}{26.39}{30}
\emline{46.40}{26.39}{31}{48.15}{26.77}{32}
\emline{48.15}{26.77}{33}{49.24}{27.80}{34}
\emline{49.24}{27.80}{35}{49.67}{30.34}{36}
\emline{49.67}{30.34}{37}{49.20}{32.19}{38}
\emline{49.20}{32.19}{39}{49.45}{33.58}{40}
\emline{49.45}{33.58}{41}{50.42}{34.50}{42}
\emline{50.42}{34.50}{43}{54.00}{35.00}{44}
\emline{54.00}{35.00}{45}{56.35}{35.18}{46}
\emline{56.35}{35.18}{47}{58.00}{35.74}{48}
\emline{58.00}{35.74}{49}{58.97}{36.66}{50}
\emline{58.97}{36.66}{51}{59.25}{37.96}{52}
\emline{59.25}{37.96}{53}{58.67}{40.00}{54}
\emline{58.67}{40.00}{55}{57.99}{41.97}{56}
\emline{57.99}{41.97}{57}{58.00}{43.52}{58}
\emline{58.00}{43.52}{59}{58.70}{44.66}{60}
\emline{58.70}{44.66}{61}{60.09}{45.37}{62}
\emline{60.09}{45.37}{63}{62.67}{45.67}{64}
\emline{48.67}{27.00}{65}{47.89}{25.00}{66}
\emline{47.89}{25.00}{67}{47.97}{23.56}{68}
\emline{47.97}{23.56}{69}{48.91}{22.66}{70}
\emline{48.91}{22.66}{71}{51.67}{22.34}{72}
\emline{51.67}{22.34}{73}{53.62}{22.58}{74}
\emline{53.62}{22.58}{75}{54.91}{22.10}{76}
\emline{54.91}{22.10}{77}{55.55}{20.90}{78}
\emline{55.55}{20.90}{79}{55.34}{18.00}{80}
\emline{55.34}{18.00}{81}{54.73}{15.78}{82}
\emline{54.73}{15.78}{83}{54.84}{14.10}{84}
\emline{54.84}{14.10}{85}{55.67}{12.98}{86}
\emline{55.67}{12.98}{87}{59.00}{12.34}{88}
\emline{59.00}{12.34}{89}{61.31}{12.21}{90}
\emline{61.31}{12.21}{91}{62.80}{11.54}{92}
\emline{62.80}{11.54}{93}{63.46}{10.31}{94}
\emline{63.46}{10.31}{95}{63.00}{7.67}{96}
\put(9.67,26.67){\makebox(0,0)[cc]{x}}
\put(66.00,45.67){\makebox(0,0)[cc]{$\overline{\text{C}}$}}
\put(66.00,8.67){\makebox(0,0)[cc]{C}}
\put(33.33,31.67){\makebox(0,0)[cc]{(NS$-$,NS$-$)}}
\emline{20.00}{26.67}{97}{21.49}{28.10}{98}
\emline{21.49}{28.10}{99}{23.05}{28.23}{100}
\emline{23.05}{28.23}{101}{25.00}{26.67}{102}
\emline{25.00}{26.67}{103}{26.48}{25.04}{104}
\emline{26.48}{25.04}{105}{27.96}{24.59}{106}
\emline{27.96}{24.59}{107}{30.33}{26.33}{108}
\emline{30.33}{26.33}{109}{31.58}{27.77}{110}
\emline{31.58}{27.77}{111}{32.83}{27.90}{112}
\emline{32.83}{27.90}{113}{34.33}{26.33}{114}
\emline{34.33}{26.33}{115}{35.25}{24.73}{116}
\emline{35.25}{24.73}{117}{36.33}{24.25}{118}
\emline{36.33}{24.25}{119}{37.58}{24.90}{120}
\emline{37.58}{24.90}{121}{39.00}{26.67}{122}
\end{picture}

\nopagebreak
Figure 3
\end{center}
The D-brane in type 0 theories couples to (NS$-$,NS$-$) closed strings and there are 
vertices in the low energy spacetime action that connect (NS$-$,NS$-$) strings to a $C$ 
and a $\overline{C}$ \cite{9811035}. 
The string diagram that contributes
to this process is a disk with $V_C$ and $V_{\overline{C}}$ operators.  These massless R-R 
vertex operators are given by
\begin{eqnarray}
V^{C_{m-1}}_i(z_i, \overline{z}_i) &=& (P_- \Gamma_{i(m)})^{AB} :V_{-1/2\,A}(p_i, z_i):\, 
:\tilde{V}_{-1/2\, B}(p_i, \tilde{z}_i):\, , \\
V^{\overline{C}_{m-1}}_i(z_i, \overline{z}_i) &=& (P_+ \Gamma_{i(m)})^{AB} 
:V_{-1/2\,A}(p_i, z_i):\, 
:\tilde{V}_{-1/2\, B}(p_i, \tilde{z}_i):\, ,
\end{eqnarray}
where we are using the notation of \cite{9603194}.  
The objects in these vertex operators are defined as
\begin{eqnarray}
V_{-1/2\, A}(p_i, z_i) &=& e^{-\Phi(z_i)/2} S_A (z_i) e^{ip_i \cdot X_L(z_i)}\, , \\
P_\pm &=& (1 \pm \gamma_{11})/2 \, , \\
\Gamma_{(n)} &=& \frac{a_n}{n!} F_{\mu_1 \ldots \mu_n} \gamma^{\mu_1} \ldots 
\gamma^{\mu_n} \, ,
\end{eqnarray}
where $S_A$ is the spin field, $\gamma_{11} = \gamma^0 \ldots \gamma^9$, and $F_n = 
dC_{n-1}$.

Under the doubling trick, the spin field $\tilde{S}_A$ will be identified as
\begin{equation}
\tilde{S}_A(\overline{z}) = {M_A}^B S_B(\overline{z})\, ,
\end{equation}
for some matrix $M$.  This matrix can be specified \cite{9603194} by considering the 
following 
OPE's.
\begin{eqnarray}
\psi^\mu(z)S_A(w) &\sim& (z-w)^{-1/2} \frac{1}{\sqrt{2}} {(\gamma^\mu)_A}^B S_B(w) + 
\ldots
\label{ope1}
\\
\tilde{\psi}^\mu(\overline{z}) \tilde{S}_A(\overline{w}) &\sim& 
(\overline{z}-\overline{w})^{-1/2} \frac{1}{\sqrt{2}} {(\gamma^\mu)_A}^B 
\tilde{S}_B(\overline{w}) + \ldots
\label{ope2}
\end{eqnarray}
The doubling trick identification for $\tilde{\psi}^\mu$ is 
$\tilde{\psi}^\mu(\overline{z}) = \eta {D^\mu}_\nu \psi^\nu(\overline{z})$, where 
${D^\mu}_\nu = ({\delta^\alpha}_\beta, -{\delta^i}_j)$.  In order for (\ref{ope2}) to be 
consistent with (\ref{ope1}), $M$ must satisfy
\begin{equation}
{(\gamma^\mu)_A}^B = {D^\mu}_\nu {(M^{-1} \gamma^\nu M)_A}^B\, .
\end{equation}
This can be rewritten as $(M\gamma^\mu) = {D^\mu}_\nu (\gamma^\nu M)$ which implies that 
$M$ is of the form
\begin{equation}
M = 
\begin{cases}
a \gamma^0 \ldots \gamma^p & \text{for p+1 odd, $\eta=1$}\\
b \gamma^0 \ldots \gamma^p \gamma_{11} & \text{for p+1 even, $\eta=1$}\\
c \gamma^0 \ldots \gamma^p \gamma_{11} & \text{for p+1 odd, $\eta=-1$}\\
d \gamma^0 \ldots \gamma^p & \text{for p+1 even, $\eta=-1$}\, .
\end{cases}
\end{equation}
To fix the phases, the OPE's
\begin{eqnarray}
S_A(z) S_B(w) &\sim& (z-w)^{-5/4} C_{AB}^{-1} + \ldots\\
\tilde{S}_A(\overline{z}) \tilde{S}_B(\overline{w}) &\sim& 
(\overline{z}-\overline{w})^{-5/4} C_{AB}^{-1} + \ldots
\end{eqnarray}
are used to find that $M^{-1} = C^{-1} M^T C$.  Since all the $\gamma^\mu$ and 
$\gamma_{11}$ anticommute with $C$, we find the phases up to an overall sign:
\begin{equation}
M =
\begin{cases}
\pm i \gamma^0 \ldots \gamma^p & \text{for p+1 odd, $\eta=1$}\\
\pm  \gamma^0 \ldots \gamma^p \gamma_{11} & \text{for p+1 even, $\eta=1$}\\
\pm \gamma^0 \ldots \gamma^p \gamma_{11} & \text{for p+1 odd, $\eta=-1$}\\
\pm i \gamma^0 \ldots \gamma^p & \text{for p+1 even, $\eta=-1$}\, .
\end{cases}
\label{mdef}
\end{equation}
From now on, we will write $M$ as $M_\eta$ to distinguish between the two forms it takes 
for fixed p.  Equation (\ref{mdef}) gives the relationships between $M_+$ and $M_-$ as
\begin{eqnarray}
M_- = \pm i M_+ \gamma_{11}\, .
\label{mrelation}
\end{eqnarray}

The amplitude for $C \rightarrow \overline{C}$ scattering off a Dp+ is \cite{9901085}
\begin{align}
A(C,\overline{C})_+ = &- \frac{i\kappa^2T_p}{2} [\frac{1}{2}\text{Tr}(P_-
\Gamma_{1(m)} M_+
\gamma^\mu)\text{Tr} (P_+ \Gamma_{2(n)} M_+\gamma_\mu) B(-t/2+1/2, -2s)\nonumber\\
&-\text{Tr}(P_- \Gamma_{1(m)} C^{-1} \Gamma^T_{2(n)} C) B(-t/2-1/2, -2s+1) \nonumber\\
&-\text{Tr}(P_- \Gamma_{1(m)}M_+\Gamma_{2(n)}M_+) B(-t/2+1/2, -2s+1)]\, .
\label{ccamp}
\end{align}
Since the Euler beta function is defined as
\begin{equation}
B(a,b) = \int_0^1 dy\, y^{a-1}(1-y)^{b-1}\, ,
\end{equation}
we see that the poles in the t channel are $m^2 = (4n-2)/\alpha^\prime$ for 
$n=0,1,\ldots$.  These poles correspond to the masses of the closed strings in the 
(NS$-$,NS$-$) sector.

To obtain $A(C,\overline{C})_-$, the amplitude for $C\rightarrow 
\overline{C}$ scattering off a Dp$-$, from $A(C,\overline{C})_+$, we must replace $M_+$ 
with $M_-$ and $e^{-\Phi(\overline{z})/2}$ with $e^{-\Phi(\overline{z})/2 - i\pi/2}$.  It 
is 
simple to check that the amplitude is invariant under replacing $M_+$ with $M_-$.  In the 
correlation function, there are two factors of $e^{-\Phi(\overline{z})/2}$ coming from the 
two R-R vertex operators.  After replacing them with $e^{-\Phi(\overline{z})/2 - i\pi/2}$, 
each one contributes a factor of $i$ for a total phase of $-1$.  In summary, we find that
\begin{equation}
A(C, \overline{C})_+ = - A(C,\overline{C})_-\, .
\end{equation}
This shows that the Dp+ and Dp$-$ couple with opposite signs to all (NS$-$,NS$-$) fields.

How can this phenomenon be understood in a direct manner?  Consider the tadpole amplitude
for emission of a closed string from a D-brane.  If the closed string is in one of the
NS-NS sectors, the amplitude is a disk with the closed string vertex operator in the
($-1$,$-1$) picture.  For a NS-NS string, the amplitude for emission from a Dp+ can be
converted into an amplitude for emission from a Dp$-$ by multiplying by $-$1 for each
factor of $e^{-\tilde{\Phi}}$ and $\tilde{\psi}^\mu$.  In the ($-1$,$-1$) picture, the
NS-NS vertex operator has as many $\tilde{\psi}$'s as does the corresponding Fock state.  
Therefore, the Dp$-$ amplitude differs from the Dp+ amplitude by a factor of
$(-1)^{\tilde{F}}$, where $\tilde{F}$ is the right-moving worldsheet fermion number of
the NS-NS closed string state.  In other words, Dp+ and Dp$-$ have the same tadpole
couplings to all (NS+,NS+) fields and opposite tadpole couplings to all (NS$-$,NS$-$)
fields.

It is clear how to generalize this to a general disk amplitude on a D-brane.  To convert
a general disk amplitude for a D+ into the same amplitude with a D$-$, we multiply by
$-$1 for each $e^{-\tilde{\Phi}}$ and $\tilde{\psi}^\mu$, and we replace $M_+$ with $\pm
i M_+ \gamma_{11}$ for each spin field.  Since a fermionic state can not transform into a
bosonic one, the number of $M_\eta$'s will be even in any nonzero amplitude, so the sign
ambiguity in that replacement is insignificant.


\section{Summary}

We set out to find the descent relations for the type 0 theories.  We found that we must
start with either a D+$\overline{\text{D}}$+ pair or a D$-\overline{\text{D}}-$ pair and
that the + and $-$ are invariant under the orbifold and kink operations.  This means we
have two copies of the usual descent relation chain for the type 0 theories: one for D+
branes and one for D$-$ branes.  We then asked why we should care about the distinction
between a D+ brane and a D$-$ brane.  While it is fairly well known that the stable D+ and
D$-$ have the same coupling to half of the massless R-R fields and equal and opposite
couplings to the other half, we have shown that the D+ and D$-$ have the same tadpole
couplings to half of the NS-NS fields and equal and opposite tadpole couplings to the
other half.\\
\vskip 10pt
\noindent\textbf{Acknowledgements:} I would like to thank A. Strominger for his comments 
on the manuscript.  This paper is based upon work supported in part by a National 
Science Foundation Graduate Fellowship.  This work was also supported in part by DOE 
Award DE-FG02-91ER40654.


\section*{Appendix A: Open String Spectrum}

In this appendix, we will find the open string spectrum on type II and type 0 D-branes.
We begin by considering the closed 
string exchange amplitudes between boundary states, which are given in \cite{9701137}.  
Motivated by the usual worldsheet duality of the cylinder diagram, this result can be 
converted into an open string loop amplitude.  The results are as follows:
\begin{equation}
\begin{array}{lcl}
\int dl \: {}_{\text{NS-NS}}\langle Bp,\eta | e^{-lH_{\text{closed}}} 
|Bp,\eta\rangle_{\text{NS-NS}}
&=& \int \frac{dt}{2t}\text{Tr}_{\text{NS}} [e^{-tH_{\text{open}}}] \\
\int dl \: {}_{\text{NS-NS}}\langle Bp,\eta | e^{-lH_{\text{closed}}} 
|Bp,-\eta\rangle_{\text{NS-NS}}
&=& -\int \frac{dt}{2t}\text{Tr}_{\text{R}} [e^{-tH_{\text{open}}}] \\
\int dl \: {}_{\text{R-R}}\langle Bp,\eta | e^{-lH_{\text{closed}}} 
|Bp,\eta\rangle_{\text{R-R}}
&=& \int \frac{dt}{2t}\text{Tr}_{\text{NS}} [(-1)^F e^{-tH_{\text{open}}}] \\
\int dl \: {}_{\text{R-R}}\langle Bp,\eta | e^{-lH_{\text{closed}}} 
|Bp,-\eta\rangle_{\text{R-R}}
&=& -\int \frac{dt}{2t}\text{Tr}_{\text{R}} [(-1)^F e^{-tH_{\text{open}}}]
\end{array}
\label{closedtoopen}
\end{equation}

We will combine equations (\ref{closedtoopen}) with the expressions \cite{0005029} for the 
type II D-branes in terms of boundary states,
\begin{equation}
\begin{array}.{rcl}\}
|Dp\rangle &=& (|Bp,+\rangle_{\text{NS-NS}} - |Bp,-\rangle_{\text{NS-NS}})\\
&& \qquad + 
(|Bp,+\rangle_{\text{R-R}} + |Bp,-\rangle_{\text{R-R}})\\
|\overline{D}p\rangle &=& (|Bp,+\rangle_{\text{NS-NS}} - |Bp,-\rangle_{\text{NS-NS}})\\
&& \qquad - 
(|Bp,+\rangle_{\text{R-R}} + |Bp,-\rangle_{\text{R-R}})\\
\end{array}
\; \text{for p even (odd) in IIA (IIB)}
\label{IIstabdstates}
\end{equation}
\begin{equation}
\begin{array}.{rcl}\}
|\widehat{Dp}\rangle &=& |Bp,+\rangle_{\text{NS-NS}} - 
|Bp,-\rangle_{\text{NS-NS}}
\end{array}
\; \text{for all p in IIA and IIB}
\label{IIunstabdstates}   
\end{equation}
and the expressions for the type 0 D-branes in terms of boundary states,
\begin{equation}
\begin{array}.{rcl}\}
|Dp,+\rangle &=& |Bp,+\rangle_{\text{NS-NS}} + |Bp,+\rangle_{\text{R-R}}\\
|Dp,-\rangle &=& |Bp,-\rangle_{\text{NS-NS}} + |Bp,-\rangle_{\text{R-R}}\\
|\overline{D}p,+\rangle &=& |Bp,+\rangle_{\text{NS-NS}} - |Bp,+\rangle_{\text{R-R}}\\
|\overline{D}p,-\rangle &=& |Bp,-\rangle_{\text{NS-NS}} - |Bp,-\rangle_{\text{R-R}}   
\end{array}
\; \text{for p even (odd) in 0A (0B)}
\label{0stabdstates}
\end{equation}
\begin{equation}
\begin{array}.{rcl}\}
|\widehat{Dp},+\rangle &=& |Bp,+\rangle_{\text{NS-NS}}\\
|\widehat{Dp},-\rangle &=& |Bp,-\rangle_{\text{NS-NS}}
\end{array}
\; \text{for all p in 0A and 0B}\, .
\label{0unstabdstates}   
\end{equation}
It is impossible for a R-R string to spontaneously convert into a NS-NS string, or vice 
versa, so we know that
\begin{equation}
{}_{\text{NS-NS}}\langle Bp, \eta^\prime | e^{-lH_{\text{closed}}} | Bp, 
\eta\rangle_{\text{R-R}} = 0\, .
\end{equation}
Now, to find the spectrum on open strings beginning and ending on a stable Dp+ in the 
type 0 theories, 
we will rewrite the closed string exchange diagram as a trace over open string states.  We 
have everything we need to perform this calculation; combining equations 
(\ref{closedtoopen}) and (\ref{0stabdstates}), we find
\begin{eqnarray}
&& \int dl\, \langle Dp, +| e^{-lH_{\text{closed}}} | Dp, +\rangle \nonumber \\
&=& \int dl\: {}_{\text{NS-NS}}\langle Bp, +|e^{-lH_{\text{closed}}} |Bp, 
+\rangle_{\text{NS-NS}} + \int dl \: {}_{\text{R-R}}\langle Bp, +|e^{-lH_{\text{closed}}} 
|Bp, +\rangle_{\text{R-R}} \nonumber \\
&=& \int \frac{dt}{2t} \text{Tr}_{\text{NS}} [e^{-tH_{\text{open}}}] + \int \frac{dt}{2t} 
\text{Tr}_{\text{NS}} [(-1)^F e^{-tH_{\text{open}}}] \nonumber \\
&=& \int \frac{dt}{2t} \text{Tr}_{\text{NS}} [(1+(-1)^F)e^{-tH_{\text{open}}}] \nonumber 
\\
&=& \int \frac{dt}{t} \text{Tr}_{\text{NS+}} [e^{-tH_{\text{open}}}] \, .
\end{eqnarray}
So we see that the open strings beginning and ending on a stable Dp+ in the type 0 
theories are NS+.  Proceeding in this manner, we can find the spectrum of open strings on 
all possible combinations of D-branes in the type 0 and type II theories.  The full 
results for the type 0 theories are given in tables 1 and 2 in section 2.  The results for 
the type II theories are given in tables 4 and 5 below.
\begin{center}
\begin{tabular}{|c|c|c|}
\hline
\multicolumn{3}{|c|}{Open Spectrum on Stable D-branes}\\
\multicolumn{3}{|c|}{(p odd in IIB, p even in IIA)}\\
\hline\hline
$\sigma=0$ & $\sigma=\pi$ & Spectrum \\
\hline
Dp & Dp & NS+, R$-$ \\
\hline
Dp & $\overline{\text{D}}$p & NS$-$, R+ \\
\hline
\end{tabular}\\
\vspace{5pt}
Table 4: The other two cases obtained by the following \\
operation under which the spectrum is invariant: $\text{D} \leftrightarrow 
\overline{\text{D}}$.
\end{center}
\begin{center}
\begin{tabular}{|c|c|c|}
\hline
\multicolumn{3}{|c|}{Open Spectrum on Unstable D-branes}\\
\multicolumn{3}{|c|}{(all p in IIA and IIB)}\\
\hline\hline
$\sigma=0$ & $\sigma=\pi$ & Spectrum \\
\hline
$\widehat{\text{Dp}}$ & $\widehat{\text{Dp}}$ & NS+, NS$-$, R+, R$-$ \\
\hline
\end{tabular}\\
\vspace{5pt}   
Table 5
\end{center}


\section*{Appendix B: Orbifold of 0A/0B}

The action of $(-1)^{F_L^s}$ can be represented as a $2\pi$ spacetime rotation on the
left-movers.  Under this rotation, the left-sector bosons (NS) are invariant and the
left-sector fermions (R) pick up a minus sign.  We can pick any spatial plane for this
rotation and for our purposes here we select the 8-9 plane.

The situation is greatly simplified if we use complexified coordinates \cite{polchII} for 
those
left-moving fields whose indices are in the 8-9 plane, 
\begin{eqnarray}
\Psi^4 &=& \frac{1}{\sqrt{2}}(\psi^8 + i \psi^9) \, , \nonumber \\
\Psi^{\overline{4}} &=& \frac{1}{\sqrt{2}}(\psi^8 - i \psi^9) \, , \\
\partial Z^4 &=& \frac{1}{\sqrt{2}}(\partial X^8 + i \partial X^9) \, , \nonumber \\
\partial Z^{\overline{4}} &=& \frac{1}{\sqrt{2}}(\partial X^8 - i \partial X^9)\, .
\end{eqnarray}
With this notation, a rotation on the left-movers by angle $\theta$ in the 8-9 plane has 
the following action on the fields:
\begin{eqnarray}
\Psi^4 &\rightarrow& e^{i\theta} \Psi^4 \, , \nonumber \\
\Psi^{\overline{4}} &\rightarrow& e^{-i\theta} \Psi^{\overline{4}} \, , 
\label{psirot} \\
\partial Z^4 &\rightarrow& e^{i\theta} \partial Z^4 \, , \nonumber \\
\partial Z^{\overline{4}} &\rightarrow& e^{-i\theta} \partial Z^{\overline{4}} \, .
\end{eqnarray}

We wish to find the orbifold of type 0A by $(-1)^{F_L^s}$.  This is an asymmetric, abelian 
orbifold with group elements \{1, $(-1)^{F_L^s}$\}.  The untwisted sector, corresponding 
to the identity element, is simply the projection of 0A on states invariant under 
$(-1)^{F_L^s}$.  It is clear that the invariant states are those in the 
sectors (NS+,NS+) and (NS$-$,NS$-$).  Let us check that we get the same result by 
representing $(-1)^{F_L^s}$ as a rotation by $2\pi$ on the left-movers.  On the NS sector 
ground state vertex operator, $1 \rightarrow 1$; the NS sector is invariant.  To consider 
the action on the R sector ground state vertex operator, we must bosonize the complexified 
fermions as
\begin{eqnarray}
\Psi^4 = e^{i H^4} \nonumber \, , \\
\Psi^{\overline{4}} = e^{-iH^4} \, ,
\end{eqnarray}
and likewise for the other fermions.
In terms of these bosonic $H$ fields, the spin operator takes the form
\begin{equation}
\Theta_s = e^{i \sum\limits_{a=1}^4 s_a H^a}\, ,
\end{equation}
where the $s_a = \pm 1/2$.
Since $\Psi^4$ transforms under the $\theta = 2\pi$ rotation as (\ref{psirot}), 
$\exp(\frac{1}{2} i H^4)$ transforms as
\begin{equation}
e^{\frac{1}{2} i H^4} \rightarrow e^{i\pi} e^{\frac{1}{2}iH^4} = - e^{\frac{1}{2}iH^4}\, .
\end{equation}
Therefore, the spin field, and subsequently the left-moving R sector vertex operator, 
picks up a minus sign from the $2\pi$ rotation; the (R+,R$-$) and (R$-$,R+) sectors are 
projected out.

In the twisted sector, the boundary conditions on the $\partial Z^4$ and $\Psi^4$ fields 
are as follows:
\begin{eqnarray}
\partial Z^4(\sigma + 2\pi) &=& e^{2\pi i} \partial Z^4(\sigma) \, , \nonumber \\
\partial Z^{\overline{4}}(\sigma+2\pi) &=& e^{-2\pi i} \partial Z^{\overline{4}}(\sigma) 
\, , \\
\Psi^4(\sigma+2\pi) &=& e^{2\pi i (\beta +\nu)} \Psi^4(\sigma) \, , \nonumber \\
\Psi^{\overline{4}}(\sigma+2\pi) &=& e^{-2\pi i (\beta +\nu)}\Psi^{\overline{4}}(\sigma) 
\, ,
\end{eqnarray}
where $\nu=0$ for R, $\nu=1/2$ for NS, and $\beta = 1$.
At first glance, it appears as though the boundary conditions are unchanged.  However, if 
we continuously change the boundary condition factor $\exp(2\pi i \beta)$ from $\beta=0$ 
to $\beta=1$, we see that the moding of the Fourier coefficients has changed from $n$ for 
both $\partial Z^4$ and $\partial Z^{\overline{4}}$ and $n+\nu$ for both $\Psi^4$ and 
$\Psi^{\overline{4}}$ to 
\begin{eqnarray}
\alpha^4 &:& n+1 \, , \nonumber \\
\alpha^{\overline{4}} &:& n-1 \, , \nonumber \\
\Psi^4 &:& n+1+\nu \, , \\
\Psi^{\overline{4}} &:& n-1-\nu \, . \nonumber
\end{eqnarray}
This phenomenon, known as spectral flow, has an important consequence for the ground state 
of the theory.  When we began with $\beta=0$, the ground state was defined as
\begin{equation}
\Psi^4_{n+\nu} |0\rangle = \Psi^{\overline{4}}_{n+1-\nu}|0\rangle = 0 \quad \text{ for } 
n=0, 1, \ldots \, ,
\end{equation}
with similar equations for the other $\Psi$.
The effect of continuously changing $\beta$ from 0 to 1 is that we replace $\nu$ with 
$\nu+1$ in these equations.  The ground state now satisfies the conditions
\begin{equation}
\Psi^4_{n+\nu+1} |0\rangle = \Psi^{\overline{4}}_{n-\nu}|0\rangle = 0 \quad \text{ for }  
n=0, 1, \ldots \, .
\end{equation}
The $|0\rangle$ state is no longer the ground state because $\Psi^4_\nu |0\rangle \neq 0$ 
and $\Psi^{\overline{4}}_{-\nu} |0\rangle = 0$.  The true ground state is
\begin{equation}
|0\rangle^\prime = \Psi^4_\nu|0\rangle
\end{equation}
since 
\begin{equation}
\Psi^4_\nu |0\rangle^\prime = \Psi^4_\nu \Psi^4_\nu |0\rangle = 0
\end{equation}
and
\begin{equation}
\Psi^{\overline{4}}_{-\nu} |0\rangle^\prime = \Psi^{\overline{4}}_{-\nu} \Psi^4_\nu 
|0\rangle = 
\{\Psi^{\overline{4}}_{-\nu}, \Psi^4_\nu\} |0\rangle = |0\rangle \neq 0 \, .
\end{equation}
However, now the GSO condition on the left-movers,
\begin{equation}
(-1)^F |0\rangle = \pm |0\rangle
\end{equation}
has become
\begin{equation}
(-1)^F|0\rangle^\prime = -\Psi^4_\nu (-1)^F |0\rangle = \mp |0\rangle^\prime\, .
\end{equation}
We see that the GSO conditions on the left-movers has been reversed.

This leaves us with the following twisted sector:
\begin{equation}   
\begin{array}{cccc}
\text{(NS$-$,NS+)}&\text{(NS+,NS$-$)}&\text{(R$-$,R$-$)}&\text{(R+,R+)}\, .   
\end{array}
\end{equation}
Of these four groups of states, we keep only those that will combine with the untwisted
sector to give us a modular invariant theory.  For abelian orbifolds, the correct criteria
for the twisted states to ensure modular invariance is level matching.  In the (NS$-$,NS+)
and (NS+,NS$-$) sectors, there is no way to obtain $L_0 = \tilde{L}_0$, so we drop these
states.

In the end, we are left with the (NS+,NS+) and (NS$-$,NS$-$) states from the untwisted 
sector and the (R$-$,R$-$) and (R+,R+) states from the twisted sector.  Combined, these 
give the spectrum of the type 0B theory as given in (\ref{0spec}).  The argument works in 
the same way to get type 0A from 0B.


\end{document}